%% file: main.tex
\documentclass[sigconf,nonacm]{acmart}
\usepackage{tikz}
\AtBeginDocument{%
  \providecommand\BibTeX{{%
    \normalfont B\kern-0.5em{\scshape i\kern-0.25em b}\kern-0.8em\TeX}}}

\makeatletter
\def\@ACM@checkaffil{%
    \if@ACM@instpresent\else
    \ClassWarningNoLine{\@classname}{No institution present for an affiliation}%
    \fi
    \if@ACM@citypresent\else
    \ClassWarningNoLine{\@classname}{No city present for an affiliation}%
    \fi
    \if@ACM@countrypresent\else
        \ClassWarningNoLine{\@classname}{No country present for an affiliation}%
    \fi
}
\makeatother

\begin{document}

\title{S3C2 Summit 2025-07: \\ Government Secure Supply Chain Summit}

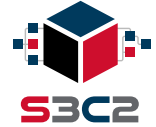
\author{Sivana Hamer$^{*}$, Pat Morrison$^{*}$, William Enck$^{*}$, Yasemin Acar$^{\dagger}$, Michel Cukier$^{\ddagger}$, \\ Alexandros Kapravelos$^{*}$, Christian Kästner$^{\mathsection}$, Dominik Wermke$^{*}$,  Laurie Williams$^{*}$}

\def \authors{Sivana Hamer, Pat Morrison, William Enck, Yasemin Acar, Michel Cukier, Alexandros Kapravelos, Christian Kästner, Dominik Wermke, Laurie Williams}

\affiliation{%
    \institution{ $^*$North Carolina State University, Raleigh, NC, USA}
}
\affiliation{%
    \institution{$^\dagger$Paderborn University, Paderborn, Germany and George Washington University, DC, USA}
}
\affiliation{%
    \institution{$^\ddagger$University of Maryland, College Park, MD, USA}
}
\affiliation{%
    \institution{ $^\mathsection$Carnegie Mellon University, Pittsburgh, PA, USA}
}

\renewcommand{\shortauthors}{Secure Software Supply Chain Center (S3C2)}
\renewcommand{\shorttitle}{S3C2 Summit 2023-06: Government Secure Supply Chain Summit}

\input{abstract}


\maketitle

\begin{tikzpicture}[overlay, remember picture]
\node[anchor=north west, %
      xshift=17.5cm, %
      yshift=-2.1cm] 
     at (current page.north west) %
     {\includegraphics[width=2.1cm]{S3C2_logo.pdf}}; 
\end{tikzpicture}

\input{intro}

\input{sbom}

\input{vulnerable-dependencies}

\input{malicious-commits}

\input{build-infra}

\input{culture}

\input{llm-security}

\input{component-container}

\input{summary}

\section{Acknowledgements}
A big thank you to all Summit participants. We are very grateful for the opportunity to hear about your valuable experiences and suggestions. The Summit was organized by Laurie Williams and Pat Morrison, and recorded by Sivana Hamer.
This material is based upon work supported by the National Science Foundation Grant Nos. 2207008, 2206859, 2206865, and 2206921.
These grants support the Secure Software Supply Chain Summit (S3C2), which comprises researchers from North Carolina State University, Carnegie Mellon University, the University of Maryland, and George Washington University. 
Any opinions expressed in this material are those of the author(s) and do not necessarily reflect the views of the National Science Foundation.

\bibliographystyle{ACM-Reference-Format}
\bibliography{literature}

\appendix

\section{Initial Discussion Questions}
\label{questions}
\begin{enumerate}

\item \textbf{SBOM, VEX, and compliance.} From your perspective, how widespread is the practice of: Producing SBOMs? Consuming SBOMs?  Storing SBOMs? Sharing SBOMs? What challenges are you facing? What will/can SBOMs actually achieve? How can they be leveraged/used? Are you seeing VEX being used?  Is the security level of VEX a concern? Do you see VEX being helpful or hurtful?  In what ways? Has self-attestation come to an abrupt halt?  Do you have a sense that companies have eased up on implementing secure software development practices?
Are there any frameworks you work to adopt/comply with (e.g. SLSA, S2C2F)?

\item \textbf{Updating vulnerable dependencies.}  What are your main concerns and pain points around updating vulnerable dependencies? 
Do they have policies around when to update? 
What kind of testing or other strategies are used  before updating to a new version?  
Are CVSS, EPSS, KEV used in your decision process?
How do SBOM and SCA tools fit into the decision?

\item \textbf{Malicious commits.}  What is the government view on malicious commits? How can malicious commits be detected? What do you think signals a suspicious/malicious commit? What role does the ecosystem play in detecting malicious commits?

\item \textbf{Build Infrastructure.}  Do you monitor your build to detect tampering? Do you scan build and CI/CD scripts for vulnerabilities? Do you use reproducible builds or hermetic builds?

\item \textbf{Culture.} What changes have you made to support supply chain security?  Has the culture changed/adapted to additional security practices?  What do you think is needed for nurturing a security-benefiting culture?

\item \textbf{LLMs and Supply Chain Security.} From the perspective of government agencies you interact with, what is your perspective on the use of large language models (LLMs) such as ChatGPT as another supply chain attack vector?
How do you protect from hallucinations from LLMs?

\item \textbf{Component and Container Choice.} Do you have a process for having components and containers approved before use? For what kinds of development?  All, production, prototypes? Do you scan components and containers? Are components and containers stored in a immutable repository for use in the build process?

\item \textbf{Research.} What are research challenges in software supply chain security?

\end{enumerate}

\end{document}

%% file: abstract.tex
\begin{abstract}
  
  Software supply chains, while providing immense economic and software development value, are only as strong as their weakest link. Over the past several years, there has been an exponential increase in cyberattacks specifically targeting vulnerable links in critical software supply chains. The attacks disrupt day-to-day functioning and threaten the security of nearly everyone on the internet, from billion-dollar companies and government agencies to hobbyist open-source developers. 
  The evolving threat of software supply chain attacks has garnered interest from both the software industry and governments worldwide in improving software supply chain security. 
  
  On Thursday, July 9th, 2025, 3 researchers from the NSF-backed Secure Software Supply Chain Center (S3C2) conducted a Secure Software Supply Chain Summit with a diverse set of 12 participants from 6 US government agencies. 
  The goals of the Summit were:
  (1)~to enable sharing between participants from different industries regarding practical experiences and challenges with software supply chain security;
  (2)~to help form new collaborations;
  and 
  (3)~to learn about the challenges facing participants to inform our future research directions. 
  The summit consisted of discussions of six topics relevant to the government agencies represented, including software bill of materials (SBOMs); compliance; malicious commits; build infrastructure; culture; and large language models (LLMs) and security. 
  For each topic of discussion, we presented participants with a list of questions to spark conversation and an overview of the discussions of two industry summit held in the past year. 
  In this report, we provide a summary of the summit. 
  The initial discussion questions for each topic are provided in the appendix. 
  
\end{abstract}

%% file: intro.tex
\section{Introduction}
\label{sec:intro}
On Wednesday, July 9th, 2025, three researchers from the NSF-backed Secure Software Supply Chain Center (S3C2)\footnote{\url{https://s3c2.org/}} conducted a day-long Secure Software Supply Chain Summit with a diverse set of 12 participants from 6 US government agencies. 
The goals of the Summit were to:
  (1)~to enable sharing between participants from different industries regarding practical experiences and challenges with software supply chain security;
  (2)~to help form new collaborations;
  and 
  (3)~to learn about the challenges facing participants to inform our future research directions. 

The Summit was conducted under the Chatham House Rule~\footnote{\url{https://www.chathamhouse.org/about-us/chatham-house-rule}}, which allows all participants to use the information discussed freely. However, disclosing who was present, their affiliations, or who said what is forbidden. 
As such, this report also adheres to the Chatham House Rule, and the affiliations of the participants are not disclosed.  
Summit participants were recruited from six US government agencies invested in software security. 
Attendance was intentionally capped to create an environment that encourages candid conversations among key stakeholders. 

The Summit consisted of discussions of seven topics, decided in advance by the participants through voting.
The voting process ensured that the topics were of interest and relevant to the companies represented. 
The discussion topics included software bill of materials (SBOMs), compliance, malicious commits, build infrastructure, culture, and large language models. 
Each topic was moderated by one of the S3C2 researchers, beginning with a list of questions to spark conversation.  These questions are provided in the appendix. For each of the seven topics, this report discusses continued themes from prior summits \cite{Summit1, Summit2, Summit3, Summit4, Summit5, Summit6, Summit7, Summit8}, followed by new ideas that emerged at this particular summit.  

Three S3C2 researchers (one professor, one researcher, and one PhD student) took notes on the discussion. 
The PhD student created a first draft of this report based on these notes, which the professor and researcher then reviewed and revised. 
The draft was then reviewed by the Summit participants and the other authors of this report, who are also S3C2 researchers and experts in software supply chain security. 

The remaining sections of this report summarize the Secure Software Supply Chain Summit.

%% file: sbom.tex
\section{Software Bill of Materials, Vulnerability Exploitability eXchange, and Compliance}
\label{sec:sbom}

The first topic of discussion was Software Bill of Materials (SBOM),  Vulnerability Exploitability eXchange, and Compliance.
An SBOM is a nested inventory of ``ingredients'' that make up a software component or product, helping to identify and track third-party components within a software system.
The Executive Order (EO) 14028~\cite{EO} states that any company that sells software to the federal government must issue a complete SBOM that complies with the National Telecommunications and Information Administration (NTIA) Minimal Elements~\cite{NtiA2025}.
Compliance standards such as the US National Institute of Standards and Technology (NIST) Secure Software Development Framework (SSDF)~\cite{SSDF} define secure software development practices.
The EO requires companies supplying software to the federal government to attest that the software is securely developed by completing a self-attestation form~\cite{Attestation}.
The European Union (EU) has also introduced the Cyber Resilience Act (CRA)~\cite{cra}, which imposes mandatory cybersecurity requirements on software manufacturers.

\subsection{Continued Themes}
\label{ssec:sbom-checkbox}

\paragraph{No Carrots and Few Sticks}
There was a sentiment among participants that they were unsure of how to promote more secure software development practices without regulations.
Similarly, prior industry summits have noted that requirements, such as SBOM production, have become mere checkmarks, and companies are providing only the minimum for compliance~\cite{Summit8}.
From a government perspective, they mentioned that going outside of what is mandated opens risks of legal cases against the government.
Hence, agencies have resorted to verifying the security of the software supply chain using their own resources.
One participant mentioned not knowing where to specifically go for help when trying to ask about requirements or pushback.
Additionally, the quality of the data contained in SBOMs is not yet sufficient.
However, there is also a lack of industry interest, as compliance is often viewed as a problem.
Still, without such cooperation, they cannot yet leverage technological advances.
Although several regulations were mentioned that might help certain supply chain practices, there was no central directive to enforce.

\paragraph{Trusting SBOMs}
Participants mentioned challenges with trusting SBOMs. 
They mentioned in 2021 that there were no tools for SBOMs.
Meanwhile, in 2025, tools are on the cusp of producing good SBOMs, but you need to work to get good SBOMs from the tools.
Still, participants discussed whether the tools' outputs could be trusted and whether the processes used were secure.
A participant mentioned that, to verify, they have requested the uncompiled code.
Then, they test and evaluate the code to produce an SBOM of code that was built and signed through an Integrated Developer Environment (IDE) plugin.
Thus, from their point of view, the SBOM is as trustworthy as it can get.
Participants also mentioned the value of how trust is offloaded in containers to Chainguard with the Wolfi distribution.

\paragraph{Producing SBOMs}
In line with prior summits, participants mentioned challenges with the production and the data within SBOMs.
There was mention of new BOMs such as HBOMs (hardware BOMS) and FBOMs (firmware BOMs), with XBOMs being the umbrella term for all types of BOMs.
There is, however, a need for a centralized supply chain format.
Furthermore, tools continue to produce different SBOMs for the same software.
There also needs to be a way to say that something is redacted, stating known unknowns.
While extending SBOMs to include custom, rich data with context that also maps to associated practices was mentioned as desirable, manually implementing this is not feasible.
They mentioned a need for ways to digitally exchange a certain amount of data, as formats like CycloneDX are verbose. 
A minimum BOM with enough pedigree is needed.

\paragraph{Consuming SBOMs}
Following prior summits, participants also mentioned challenges in consuming SBOMs.
SBOMs, by themselves, merely provide data.
Yet, other sources are needed to transform them into actionable insights and intelligence. 
Additionally, a participant mentioned that the quality of the data and the tools has been, as a consumer, frustrating.
Still, consuming SBOMs is necessary, as it is impossible to produce a comprehensive theoretical landscape without them.
Furthermore, with all SBOMs produced, they could focus on defending systems rather than on generating SBOMs.
SBOMs are typically generated for an individual component and so they lack both detail on transitive dependencies of the component and visibility across the multiple components of which systems are often composed. 
Tooling support to address each of these concerns would help practitioners assess their status and risk more completely.  
In terms of the benefits of SBOM consumption, one participant remarked: ``With the use of SBOMs, a full  Source Code Analysis (SCA) is not needed every time a new version of something comes out. Using the SBOM to find just the components that need to change seemed to be a big help.'' 
Another participant remarked that ``this kind of visibility [SBOMs] will lead to companies being embarrassed about using very out-of-date or vulnerable software and will increase security updates accordingly to avoid this.

\paragraph{Low VEX Adoption}
Vulnerability Exploitability eXchange (VEX) adoption has remained low. 
The participants mentioned that VEX originated from a large software supplier attempting to reduce customer support costs, specifically identifying vulnerabilities they did not want to address. 
Furthermore, sharing information about vulnerabilities can expose organizations to legal liability.
For example, they mentioned the case of the US Securities and Exchange Commission (SEC) against SolarWinds~\cite{SECvSolarWinds2023}.
Companies are not generating VEX statements without being instructed to do so by organizations.

To determine whether vulnerabilities are present and reachable, participants employed a multiple approach that utilized open-source databases, paid third-party services, and internal threat intelligence. Examples of data sources mentioned include Open Source Vulnerability (OSV), GitHub Advisory Database, vulndb, and the CISA Known Exploitable Vulnerability (KEV) list. Some companies pay vendors or third parties to map vulnerabilities in their products (to assets).

\subsection{New Ideas}

\paragraph{Shifting US Regulations}
Participants discussed how regulations are shifting within the US.
In 2021, EO 14028 mandated the issuance of SBOMs for companies that sell software to the US government~\cite{EO}.
In 2022, Memorandum M-22-18~\cite{EO_memo} issued guidance that the set of practices for secure software development is within NIST SSDF~\cite{SSDF}.
In 2023, the US CISA issued the self-attestation form~\cite{Attestation}, to which company CEOs attest to practices derived from NIST SSDF. 
In 2025, a new EO 14144~\cite{EO14144Cybersecurity} was issued, expanding attestation and storage.
Notably, attestations were expanded from the self-attestation form~\cite{Attestation}, which had more flexibility (as CEOs attested to the best of their ability), and requirements to self-attest could be waived (stating that software was secure as it was built securely).
Participants stated, as an analogy, that one can have a clean kitchen and also make dangerous food.
Instead, EO 14144 mandated technical, machine-readable attestations that tracked provenance.
However, the attestation of EO 14144 was walked back as it had not yet been thoroughly thought out, as stated in a later EO 14306~\cite{EO14306Ammendment}.
Still, the self-attestation form from 2023 is in effect.

\paragraph{Mapping Attacks to Frameworks}
Extending the need to transform SBOMs to actionable insights and intelligence mentioned in Section~\ref{sec:sbom}, participants discussed mapping information from software supply chain security frameworks to attacks, namely, through MITRE Adversarial Tactics, Techniques, and
Common Knowledge (MITRE ATT\&CK)~\cite{mitreATTACK} and MITRE Shield~\cite{mitreSHIELD}.
Specifically, a participant mentioned leveraging Artificial Intelligence (AI) by utilizing lexicons of terms across frameworks to help identify regulatory violations.
We discussed the research we have done mapping Cyber Threat Intelligence (CTI) reports of attacks to MITRE ATT\&CK and later to software supply chain security frameworks~\cite{hamer2025closing}.

\paragraph{Quantum Supply Chains}
There was a mention of security concerns for quantum supply chains.
Quantum supply chains focus on the supply chain needs, from collecting minerals to construct the system, to the final software deployed in quantum systems.
Quantum systems can also be used to attack traditional software supply chains, leveraging their computational speeds.
The participant stated that the environment has not yet been secured.

\paragraph{Reporting Gaps}
A participant stated that there are multi-year gaps between reporting Common Vulnerabilities and Exposures (CVEs) to the National Vulnerability Database (NVD)~\cite{nistNVD} and patching. 
CVE Numbering Authorities (CNA) are taking several months to declare the CVEs.
Thus, software is being adopted despite known vulnerabilities after several months.
Patches are then applied too late within the developer lifecycle, sometimes after the software is already end-of-life.
Thus, tools fail to detect vulnerabilities during scanning.
At the same time, a participant noted that the Common Vulnerability Scoring System (CVSS) scores do not need to be provided by the NVD, as a mature organization can generate its own CVSS scores.
Therefore, timely feedback is necessary to reduce gaps, and central efforts are needed to mitigate vulnerabilities across agencies.

\paragraph{Vendors Pushback on SBOMs}
A participant mentioned that they are still getting pushback on requesting SBOMs from US government agencies.
In contrast to industry summits, the government requires SBOMs from vendors, notably during contract negotiations.
Vendors are beating around the bush, saying they cannot do it, providing partial SBOMs, and claiming that Open Source Software (OSS) is used without specifying specific dependencies.
Namely, the vendors stated that in the absence of a direct government requirement, agencies cannot force developers to provide a full SBOM.
As such, they were wondering if there is any language or directive they could use during such contentions, which leaves major security issues for them as dependents.
They considered SBOMs as the first level needed to verify if the software may be compromised, without requiring access to the software itself.
Other participants mentioned that they were horrified by the pushback.
They also agreed that the SBOM language is voluntary and not yet a direct White House requirement with a standard SBOM format.

Participants then discussed the strategies to tackle such pushback.
Although it is possible to take a legal route through a contract, it is a waste of taxpayer money.
A participant mentioned that leveraging the size of their government agency has been effective.
Yet, not all agencies have such a size, and there are not always alternatives to software from vendors.
They also stated that the information-sharing requirements from the Department of Homeland Security (DHS) or the Evidentiary Act could be helpful.
A participant also mentioned using counterintelligence to study the behavior of vendor software. 
Hence, while they may claim they have no vulnerabilities, they demonstrate that the software they are using exhibits certain behaviors. 
Hence, they ask the vendor to either provide SBOMs for confirmation or allow a simulation. 
Additionally, they mentioned that during the summer, the Cybersecurity and Infrastructure Security Agency (CISA) would update the Minimum Elements for a Software Bill of Materials (SBOM)~\cite{cisa2025elements}, building on the National Telecommunications and Information Administration (NTIA) elements from 2021.

\paragraph{Continuous SBOMs}
Participants mention the need to go from a static SBOM to a continuous SBOM.
SBOMs could thus be performed at delivery to continuously search for bad ingredients.
The process should be automated through Application Programming Interfaces (APIs) to reduce friction.
Although data size could be a problem, metadata deltas could be sent.
Others mention that rather than updating the SBOM, a new one would be generated, and different SBOMs could be produced throughout the development lifecycle.
Hence, continuous risk monitoring and pooling could be guaranteed.
Such SBOMs could be enforced through contracts to have guarantees.
Still, such a vision seemed like a dream for participants.

%% file: vulnerable-dependencies.tex
\section{Updating Vulnerable Dependencies}
\label{sec:vulnerable-dependencies}

Modern software relies on dependencies as building blocks, allowing for rapid reuse and lower upfront development costs. However, dependencies also have drawbacks, namely dependency management. In particular, keeping up with dependency vulnerability patches can be overwhelming and requires significant manual effort from already overburdened developers. It can be difficult to determine which vulnerabilities require time to address and which do not, leading to what some refer to as \textit{patch fatigue}.

\subsection{Continued Themes}

\paragraph{Alert Fatigue}
Participants agreed with the analogy that the output of tools is similar to a `firehose' dispensing a large amount of information to users.
Still, it was noted that there are perverse incentives for tools to produce large amounts of data, thereby creating false positives, as tools want to appear valuable.
Participants mentioned the need for different visualizations of a given dataset customized to the concerns of different roles.
Participants also mentioned that the overwhelming number of notifications makes users feel unsure if they are being hacked.
A good consumer might not click on any information to avoid being phished, as they may be unsure if the information is legitimate.
Hence, trust from the systems is declining.

\paragraph{Reachability}
The participants started by discussing the current state of reachability and that there are two stories: (i) we are doing well, and (ii) how much do you want to bet that we are doing well. 
Participants wondered if anything meaningful could be said about the tools themselves.
Furthermore, they questioned whether the exploitability of the vulnerability in the codebases should also be a target.
Given that participants also expressed that tools can also be non-deterministic, more research into the effectiveness of reachability analysis tools is needed.

\paragraph{Patching}
Participants discussed different strategies for patching software.
Specifically, participants mentioned patching based on the software's criticality. 
Hardware constraints, such as old, end-of-life, and custom equipment, as well as operational constraints like the need to complete a critical mission, place limits on the appropriate triggers for applying patches to governmental systems. 
In extreme cases, hardware constraints, such as outdated or critical equipment not connected to the Internet, must be patched manually.
Additionally, scanning is limited in scope due to security concerns for critical infrastructure.
Hence, determining which patches to apply and when to apply them remains a problem with no one-size-fits-all solution.
Automated patches were mentioned, but no one was sure whether any systems used such an approach.
Participants mentioned forking, copying existing code, and making local updates as an alternative strategy. 
Increasing technical debt and managing the need to converge with the original code when new features are added are recognized as costs of the strategy.

\subsection{New Ideas}

\paragraph{Triggers of Updates}
Participants began discussing what actually triggered updates. 
For classic enterprise software, visibility is limited to the software vendor, which you must contact directly.
Yet SBOMs, in theory, enable continuous analysis of software to identify vendors.
When contacting vendors, negotiations arise about whether the software can be updated.
Participants mentioned another source of triggering updates, Software Composition Analysis (SCA) tools, but these tools produce too many triggers.

One participant mentioned a recent refinement to vendor-driven automated patching; delaying application of a vendor update for anywhere from a day to several weeks, e.g., pnpm's 'minimumReleaseAge' parameter~\url{https://pnpm.io/blog/releases/10.16}, to allow vulnerable releases to be resolved by the vendor rather than the consumer.

\paragraph{Risk Tolerance}

Participants discussed the varying levels of operational risk for updating.
For some, as we are not at war, they wonder why take the risk of making a decision.
Additionally, it was mentioned that for some, taking such risks does not matter.
As an example outside of the software supply chain, there was mention of the Titan Submersible from OceanGate, which imploded, as a significant amount of risk was taken.
As such, they mentioned that vulnerability patching is not a concern, given the high risk tolerance.
For many, not patching does not stop day-to-day operations, so why stop developers?
Although we could generate more data, it will not matter if the consumers do not understand or care.
Furthermore, people tend to care only after an exploit has occurred and gained access to the system.
It is also not feasible to expect employees to become cyber experts, even though they are taking on cyber risks with their data.
The problem is inherently human, and if the problem were easy to solve, the industry would not have been struggling with it for decades.

%% file: malicious-commits.tex
\section{Malicious Commits}
\label{sec:malicious-commits}

Malicious commits were the next topic of discussion.
The XZ Utils backdoor (discovered in early 2024) was the result of malicious commits. The adversaries gained the trust of the XZ Utils maintainer by starting with committing to the XZ Utils library.
After a pressuring campaign and years of gaining trust as a maintainer, the malicious actor introduced the backdoor through malicious commits.
Since the XZ Utils backdoor, social engineering and malicious commits have been a frequent topic of discussion.

Participants identified a trust difference between proprietary and open-source software (OSS); for proprietary software, consumers must trust the vendor, while, for OSS software, the source code and the process that delivered the code is the locus of trust. In both the proprietary and OSS case, participants noted the importance to them of identifying who was involved in developing and providing the software and assessing the trust relationship with the parties involved. 

\subsection{Continued Themes}

\paragraph{Intent}
Participants discussed the difficulty of determining the intent behind a commit.
Specifically, it was noted that the most effective way to generate a malicious commit is to utilize the Common Weakness Enumeration (CWE) example code.
One could state it was not malicious, as it is a known weakness.
Additionally, they mentioned that current malware sensors are not particularly good, even more so considering how capabilities change over time.
A participant observed that while we might struggle with deobfuscating code, we are still good at identifying that code has been obfuscated. We may not be certain if something is malicious, but we are aware of what constitutes bad coding practices.
At the same time, most attacks are just copycats, and the real concern is zero-day vulnerabilities.

\paragraph{Zero-Trust}
A discussion was held on zero-trust models for consuming open-source software.
Participants noted an overall increase in foreign influence in the open source community. Some organizations are checking the identity of all open-source maintainers. Banning software could be an option if authorities find inappropriate foreign influence. They mentioned that this is similar to approaches used by tool monitors for voter registration.
Lastly, participants noted that without such safeguards, adversaries can easily use sock puppet accounts.

\paragraph{Signals}
Participants noted that signals could be used as heuristics.
Specifically, the mentioned tools to measure project health, such as OpenSSF Scorecard~\footnote{\url{https://github.com/ossf/scorecard}}, Community Health Analytics in Open Source Software
(CHAOOS)~\footnote{\url{https://chaoss.community/}}, and HIPCheck~\footnote{\url{https://github.com/mitre/hipcheck}}.
A participant mentioned that if an organization prioritizes maliciousness, it will be more inclined to use heuristics.
They explained as an analogy, it is similar to someone with allergies being more attentive to food labels. 
Still, heuristics need to take into account variance in behaviors.
Participants identified the importance of the criticality of the components when looking at the signals.

\paragraph{Supporting Open-Source Software}
Participants mentioned the need to support small open-source projects that may not want to adopt practices.
The projects may come from transitive dependencies of a component.
Still, they mentioned the case of how Alpha Omega was successful with Apache Airflow while finding or creating alternatives after auditing dependencies~\footnote{\url{https://www.first.org/resources/papers/vulncon25/Airflow-Beach-Cleaning-Securing-Supply-Chain-Vulncon-April-2025.pdf}}.

\subsection{New Ideas}

\paragraph{Partial-Trust}
Although some participants discussed zero-trust stances at length, others took an opposing stance.
They mentioned two different models of trust for proprietary software and open-source software.
For proprietary software, what is vetted is the company through past work relationships, the software produced, and the versions.
Hence, what is happening in the software is a black box, but the identity of the vendor is known.
Meanwhile, identity is an antithesis to the open-source community, and although no identity is known, there is still an associated reputation.
In OSS, you can see how the process was developed and consider different risk factors, such as the level of maintenance or countries associated with the maintainers.
Hence, the identity is a black box, but everything about how the software was developed is visible.

\paragraph{No Eyeballs}
Linus' law states that ``given enough eyeballs, all bugs are shallow''~\cite{raymond1999cathedral}.
Thus, if we have 100 maintainers in a project, a portion of the other 99 should be looking at the code.
In theory, if someone introduces malicious code, other developers, both within and outside the project, can review the code.
Still, a participant stated that no one is really looking at and reviewing the code.
Hence, the participant stated that the open source code needs to be either examined by who is committing or what is being committed.

%% file: build-infra.tex
\section{Build Infrastructure}
\label{sec:build-infra}

Various build platforms and CI/CD tools support developers in automating software development processes, including building, testing, and deployment. Build infrastructure enhances the integrity of software builds by creating documented and consistent build environments, isolating build processes, and generating verifiable provenance. 
Additionally, reproducible builds contribute to this integrity by making the build output deterministic and verifiable, ensuring builds can be consistently reproduced and verified.

\subsection{Continued Themes}
Participants reviewed current progress toward trusting builds and communicating that trust.

\paragraph{Tool Support}
Participants emphasized the importance of incentivizing secure dependency selection and the role of tools like languages, frameworks, and Dependabot-like solutions. One participant noted that "Adopting particular types of programming languages or frameworks can reduce a system’s risk for entire classes of vulnerabilities."

\paragraph{Isolating Builds}
Some participants mentioned success in isolating builds, with a participant noting that hermetic~\footnote{
Hermetic - When given the same input source code and product configuration, a hermetic build system always returns the same output by isolating the build from changes to the host system - \url{https://bazel.build/basics/hermeticity}} builds are being adopted.
However, other participants mentioned challenges for isolating builds. 
Notably, how does one limit access sufficiently while allowing for remote work?
There needs to be a differentiation between a nation-state actor and someone working on the beach using developer systems.
How does one also secure higher-protected systems from risk?
Zero-Trust was mentioned as a possible solution.

\paragraph{Attestation and Provenance}
Participants mention the use of attestations, particularly in-toto~\cite{torres2019toto}, during the build process to indicate that certain steps are being done.
While in-toto attestations were deemed valuable, participants indicated an end-to-end simplified in-toto attestations are needed.
One participant reported that ``some companies are just making their projects open source so that they do not need to do the attestation''. 
A participant also mentioned wanting to know what the gaps are in the in-toto implementations.
Given the set of claims made by individual attestations, there is also a need to have policy languages to express statements based on them.
Participants mentioned the Supply Chain Integrity, Transparency, and Trust (SCITT)~\footnote{\url{https://scitt.io/}} as an initiative within the space.
There is also a need to formalize identity as part of the attestation, hence providing provenance.

\subsection{New Ideas}

\paragraph{What Will Be The Panacea?} %
There was a discussion about the solution to building infrastructure attacks.
There was mention of six possible states we may be in: 
\begin{enumerate}
    \item Infrastructure is an underspecific problem. 
    Thus, we do not know what we want, much like the Supply Chain Levels for Software Artifacts (SLSA) framework is still evolving.
    \item We have a general idea of what we want, but it is so context-specific that we cannot easily characterize the problem. Hence, it is hard to map to a general function. 
    \item We know what we want and what we need is specific, but we do not yet have the tools. 
    Hence, tools would need to be assembled to provide a more complete picture.
    \item Tools exist, but they are not usable. in-toto was mentioned as an example.
    \item Tools exist and are usable, yet we cannot easily show their usability. 
    The cost of compliance is not easy to show and we need to give data or stand over the shoulder of developers. There was a specific mention of the SCITT initiative as one such solution.
    \item We are solving, or the market believes we are solving, the wrong problem.
    Hence, we will centralize trust to a company. For example, in the case of Chainguard with Wolfi~\footnote{\url{https://wolfi.dev}}.
\end{enumerate}

However, at the moment, we have evidence for being in each of the six states.
As a community, we are currently scattered.

\paragraph{Education}
Participants mentioned concerns about whether we are teaching the topics discussed.
Without education, the new workforce will have no idea what to do.
Furthermore, the workforce may believe that just using GitHub is sufficient.

%% file: culture.tex
\section{Culture}
\label{sec:culture}

Attendees discussed how they address and foster a culture of security within their organizations.
While security is often evaluated in relation to compliance with standards, participants noted that motivating security practices through compliance is a challenging task.

\subsection{Continued Themes}

\paragraph{Impacts of Regulation}
Building on what was stated in Section~\ref{sec:sbom}, participants noted that compliance has been a significant force in enhancing security posture.
Yet despite the tremendous growth, compliance is not motivating.
Still, regulation is seen as necessary, as companies state that without it, it is merely a requirement rather than something enforceable.
The adoption of SBOMs and CISA's Secure by Design was largely driven by compliance.
There was a discussion that, without regulation, managers are unsure how to ensure that change is adopted by vendors. 
Disruptive actions are thus seen as needed by the government to allow agencies to take a proactive stance against adversaries.

\paragraph{Fostering a Security Mindset}
In line with the prior summit~\cite{Summit8}, participants mentioned the need to foster a security mindset. A participant commented that ``Making the security needs more personal to each [team member] supports their internalization of how the organization’s security needs directly relate to and impact the things they care about. However, even with effective communication, building the culture will take time, and progress will be slow.''.
Different communities think in different ways, and there needs to be different ways to communicate with each.
Understanding the nuances of software development is not intuitive for other fields, as there are no tangible artifacts for software.
Thus, there needs to be a higher awareness of how systems can be exploited.
Participants also mentioned the value of Capture the Flag (CTF) exercises in helping to change people's mindset to think more like a ``criminal''.
Security must also be perceived as something that can happen to them.
Other ways to foster such a mindset include pushing back, stating that it is impossible for certain teams to produce a simple JSON for an SBOM.

\paragraph{Security Across the Entire Organization.}
Following the prior summit~\cite{Summit8}, participants discussed how security needs to be a cross-organizational effort.
Several participants called for standardizing security engineering practices, pointing out that, as current United States software regulations stand, no laws place liability on software producers when their products fail in completely benign standard environments, much less in any other context. 
Other participants disagreed, reasoning that higher expectations for producers would lead to less innovation, as they would adopt a risk-averse approach. 
Although we may know what to do, solutions are very expensive and require expertise.
Yet, communicating the difficulties of implementing solutions to key stakeholders is necessary, as it is challenging to explain economic cybersecurity arguments.
One of the most fundamental problems is that there is no good way to measure harm.
Without such understanding, cheaper versions of solutions may be adopted to lower costs, although humans should always be involved.
Awareness of the consequences of non-adoption of tasks must also be made clear.
For example, if an incident happens during a weekend, developers will be forced to patch, however painful it may be.
On the other hand, C-Suites likely do not want to appear on the front page of the news due to an attack.

\subsection{New Ideas}

\paragraph{Need for Analogies}
Participants mentioned that the Twinky analogy has been an effective strategy for communicating the value of SBOMs across organizational hierarchies and disciplines.
Another example of an analogy was with access control, where giving a key to your house to only the people you trust.
As such, participants emphasized the need for similar analogies throughout the software supply chain. 
Analogies need to be both simple and compelling.
Finally, they stated that software supply chain security is similar to asbestos, in that risk cannot be fully eliminated.

\paragraph{Need a Shared Vision}
Participants discussed how the software supply chain lacks a shared vision and priority.
As an example, participants mentioned how the US strived to have humankind on the moon, allowing for a shared sense of vision and priorities across US agencies and the public.
A shared vision thus helps, as different groups often speak different languages, and messaging can become diluted.
However, for that to happen, we need to have a moon that we can reach.
In other words, we do not know what success looks like yet for us.
How do we know if someone or all of us are done securing the software supply chain?

%% file: llm-security.tex
\section{LLMs and Security}
\label{sec:llm-security}

Recent advancements in AI technology, particularly the emergence of LLMs, have led to widespread adoption by companies and developers alike.
While this technology appears to be very useful, it is still in the early stages of its development lifecycle. 
We asked participants how they leverage the recent advances in AI and whether there are policies around the use of these tools. Participants described a wide range of use cases including generating assurance proofs, creating personas for attacks, detecting malicious commits, fixing documentation problems, fixing code readability, generating fuzzing targets, identifying potentially malicious commits, and providing code suggestions to fix bugs and vulnerabilities.

\subsection{Continuing Trends}

\paragraph{Prompt Engineering}
Participants mentioned that the models range from amazing to dopey. 
Still, with prompt engineering, you can shape models into anything.
Hence, escape from the guardrails and target certain areas.
The models can also write code to tell you what to do, or even use models to create prompts to then tell the models what you want.

\paragraph{Hallucinations}
Participants mentioned challenges with models hallucinating. 
Particularly, they want systems to provide a model of risk that bounds the uncertainty of the answers, as a way to determine the level of hallucination.
As an issue, they also mention how packages can be hallucinated by the models.
Hence, attackers could use the fake packages in attacks.

\paragraph{Poisoning}
Participants mentioned how public models can be vulnerable, even with millions of downloads.
As such, knowing what information was used to produce the model, such as through an AI SBOM, would be valuable.
As a solution, a participant also mentioned that conferences have artifact evaluation and vet papers.
Thus, a similar model may be implemented for models.
Additionally, building on package hallucination, participants mentioned how attackers could input fake packages into sites like StackOverflow to also poison models.

\paragraph{Licenses and Intellectual Property}
Participants mentioned that the datasets and training data used by the models are not public.
Hence, what the models are doing with data is obscured, and they were unsure how to work around such restrictions.
Participants mentioned concerns about how companies could violate contracts and train their models on private data.

\subsection{New Ideas}

\paragraph{AI as Software}
Prior summits have extensively discussed the potential benefits and challenges of LLMs.
Still, this is the first summit where participants stated that LLM-based systems are still software.
You can therefore create CVEs for AI products that correspond to CWEs.
As such, LLM-based systems are not new, and if you analyze such systems as AI, you are missing the bulk of the problem.

\paragraph{SBOMs for AI}
Participants noted that SBOMs for AI have also begun to emerge.
Such SBOMs understand the current landscape, to understand what is in their software and be able to improve development.
Still, as AI systems have different software licenses, data licenses, and systems used, differences need to be considered when creating new SBOM formats.
Additionally, there needs to be standards of what an SBOM for AI should include, as everybody is doing it differently.
While some creators include model and data, the operation also needs to be considered.
While there may be some temptation to put everything within these new SBOM types, it may not be prudent.
Overall, participants were not sure what the bare minimum fields should be for such systems.
Participants also expressed concerns about how SBOMs have yet to be fully implemented without full compliance, and worry that a similar situation may occur with AI companies.
There was mention of CISA's AI Tiger Team Use Cases~\footnote{\url{https://github.com/aibom-squad/SBOM-for-AI-Use-Cases}}, which provide some support for SBOMs for AI, but could be further extended to be even more powerful.

\paragraph{Secure Development of AI}
Most AI systems are being developed in Python. 
Participants considered that Python lacks secure software development.
Furthermore, as many systems are developed by data scientists, participants also expressed concerns about whether data scientists have received any training in security.
Participants commented that, because the risks of LLM use are poorly understood and highly imminent, there is a desire to keep LLMs far away from more sensitive artifacts, such as weapons systems.

\paragraph{Shifting Model Assumptions}
Participants also expressed concerns that the assumptions underlying LLM models have shifted.
Notably, the Model Context Protocol (MCP) has emerged.
Hence, the architecture of creating models has shifted and is no longer static.
Participants also mentioned Retrieval-Augmented Generation (RAG) AI models, which are trained on specific databases.
Particularly, using RAG was mentioned to have made an application much cheaper.

\paragraph{Emerging AI Frameworks}
AI frameworks have begun to emerge, providing guidance on the use of these systems.
For example, the MITRE Adversarial Threat Landscape for Artificial-Intelligence Systems (ATLAS)~\cite{mitreATLAS} was created to provide attack techniques and cases for AI systems, similar to the MITRE ATT\&CK framework.
Still, participants mention that more regulations are needed and there are no current answers, with contracts absorbing all risk if something goes wrong.
Hence, participants want the policy to know what to do and how to do it right.

\paragraph{AI Not an End All be All}
Participants mention that companies have stated that AI is going to ``fix'' everything. 
As such, the models are being deployed everywhere without many users of the products knowing.
Yet the costs are elevated, and models are not always effective.
A participant stated that although people want AI to be the solution to all their problems, what is actually wanted is automation.

\paragraph{Amnesia}
Participants mentioned that, as a challenge, not only hallucination, but also ``amnesia''.
Participants defined models as having amnesia when the model cannot recall prior conversations or even the same conversation.
They were wondering about possible solutions, although no solution was mentioned.

%% file: component-container.tex
\section{Component and Container Choice}
\label{sec:component}

Open source dependencies vary widely in quality, maintenance, origin, and licenses. Every dependency introduces value and risk, and once a dependency is incorporated into a project, it is often hard to replace. Therefore, it is important to have a policy that governs how software developers may choose new dependencies.

\subsection{New Ideas}

\paragraph{Creating a Paved Path}
Participants mentioned the need to create a paved path for others.
As examples of paved paths, they mentioned work done by Chainguard, Google, and JFrog.
Hence, we should focus on how we can front-load what is possible with a gentle path to make security more enticing.
Later on, improving the quality of what was done can be gradually done.
Still, what is required for the paved path for each company is individual, as not everyone, for example, needs to run SCA tools.
They also mentioned, as an idea, a paved path as a service.

%% file: summary.tex
\section{Executive Summary}
\label{sec:summary}

The July 2025 S3C2 Secure Software Supply Chain Government Summit brought together practitioners from 12 participants from 6 US government agencies to discuss first‑hand experiences with defending modern software supply chains. 
Seven topics were discussed: (i) SBOMs, VEX, and compliance; (ii) updating vulnerable dependencies; (iii) malicious commits; (iv) build infrastructure;  (v) culture; (vi) LLMs; and (vii) component and container choice.
The topics revealed a set of unifying themes that underscore the evolving nature of supply chain security challenges. 
All of the panels continued the discussion of themes raised in prior summits, though these discussions often highlighted new nuances and anecdotes.

The panels also revealed the continued maturation of the industry, highlighting new ideas and changes in viewpoint.
For the \emph{SBOM}, \emph{VEX}, and \emph{Compliance} panels, there was a sense of frustration that participants were unsure how to drive change without compliance.
Meanwhile, SBOMs continue to face challenges related to trust, production, and consumption, with vendors in some cases still resisting the sharing of full files.
VEX has continued to face low adoption due to concerns of legal liability.
Still, new US regulations and guidance have emerged, shifting the landscape for industry and government organizations.
The \emph{updating vulnerable dependencies} revealed that we continue to find updates remain challenging due to overwhelming notifications, a lack of reachability tools, and the absence of a standard patching method.
Yet, participants discussed new important factors that need to be considered for understanding how and why we are updating dependencies, such as the triggers for updates and human risk tolerance. 
The \emph{Malicious Commits} panel revealed that different strategies can be employed when consuming dependencies: zero-trust, where all actions are vetted; or partial-trust, where certain actions are vetted depending on the type of dependency.
Furthermore, we have yet to make significant progress on attributing intent and creating signals.
Still, actively supporting open-source software and more eyeballs checking the code are mentioned as strategies to help mitigate risk. 
The \emph{Build Infrastructure} panel revealed how progress has been made isolating builds, attesting to actions, and provenance.
Yet, as an industry, we are still uncertain about our current state or what the solution is to mitigate the build infrastructure attack vector.
More education is also needed to inform developers of best practices for the build pipeline.
The \emph{Culture} panel revealed how regulation has been a force for improving security posture, how organizations are fostering security mindsets, and demonstrated that security can be enhanced across the organization.
Additionally, it discussed solutions to improve culture outside of compliance through creating analogies and a shared vision.
The \emph{LLMs and Security} panel revealed a more mature approach to addressing the security concerns of LLM models, treating models as software.
Participants also discussed how regulations are emerging and shifting in the space.
Finally, the \emph{Component and Container Choice} panel highlighted the need to create paved paths to facilitate organizations' adoption of security practices.

%% file: main.bbl

\begin{thebibliography}{25}


\ifx \showCODEN    \undefined \def \showCODEN     #1{\unskip}     \fi
\ifx \showDOI      \undefined \def \showDOI       #1{#1}\fi
\ifx \showISBNx    \undefined \def \showISBNx     #1{\unskip}     \fi
\ifx \showISBNxiii \undefined \def \showISBNxiii  #1{\unskip}     \fi
\ifx \showISSN     \undefined \def \showISSN      #1{\unskip}     \fi
\ifx \showLCCN     \undefined \def \showLCCN      #1{\unskip}     \fi
\ifx \shownote     \undefined \def \shownote      #1{#1}          \fi
\ifx \showarticletitle \undefined \def \showarticletitle #1{#1}   \fi
\ifx \showURL      \undefined \def \showURL       {\relax}        \fi
\providecommand\bibfield[2]{#2}
\providecommand\bibinfo[2]{#2}
\providecommand\natexlab[1]{#1}
\providecommand\showeprint[2][]{arXiv:#2}

\bibitem[SEC(2023)]%
        {SECvSolarWinds2023}
 \bibinfo{year}{2023}\natexlab{}.
\newblock \bibinfo{title}{Securities and Exchange Commission v. SolarWinds Corp.}
\newblock \bibinfo{howpublished}{Docket No. 1:23-cv-09518-PAE, S.D.N.Y. (filed Oct. 30, 2023; dismissed Nov. 20, 2025)}.
\newblock
\urldef\tempurl%
\url{https://www.courtlistener.com/docket/67927585/securities-and-exchange-commission-v-solarwinds-corp/}
\showURL{%
\tempurl}


\bibitem[{Cybersecurity and Infrastructure Security Agency}(2025)]%
        {cisa2025elements}
\bibfield{author}{\bibinfo{person}{{Cybersecurity and Infrastructure Security Agency}}.} \bibinfo{year}{2025}\natexlab{}.
\newblock \bibinfo{title}{{2025 Minimum Elements for a Software Bill of Materials (SBOM)}}.
\newblock \bibinfo{howpublished}{\url{https://www.cisa.gov/resources-tools/resources/2025-minimum-elements-software-bill-materials-sbom}}.
\newblock


\bibitem[{Department of Homeland Security Cybersecurity and Infrastructure Security Agency (CISA)}(2024)]%
        {Attestation}
\bibfield{author}{\bibinfo{person}{{Department of Homeland Security Cybersecurity and Infrastructure Security Agency (CISA)}}.} \bibinfo{year}{2024}\natexlab{}.
\newblock \bibinfo{title}{Secure Software Development Attestation Form}.
\newblock
\newblock
\urldef\tempurl%
\url{https://www.cisa.gov/sites/default/files/2024-04/Self_Attestation_Common_Form_FINAL_508c.pdf}
\showURL{%
\tempurl}


\bibitem[Dunlap et~al\mbox{.}(2023)]%
        {Summit2}
\bibfield{author}{\bibinfo{person}{Trevor Dunlap}, \bibinfo{person}{Yasemin Acar}, \bibinfo{person}{Michel Cucker}, \bibinfo{person}{William Enck}, \bibinfo{person}{Alexandros Kapravelos}, \bibinfo{person}{Christian Kastner}, {and} \bibinfo{person}{Laurie Williams}.} \bibinfo{year}{February 2023}\natexlab{}.
\newblock \showarticletitle{S3C2 Summit 2023-02: Industry Secure Supply Chain Summit}.
\newblock \bibinfo{journal}{\emph{http://arxiv.org/abs/2307.16557}} (\bibinfo{year}{February 2023}).
\newblock


\bibitem[Enck et~al\mbox{.}(2023)]%
        {Summit3}
\bibfield{author}{\bibinfo{person}{William Enck}, \bibinfo{person}{Yasemin Acar}, \bibinfo{person}{Michel Cucker}, \bibinfo{person}{Alexandros Kapravelos}, \bibinfo{person}{Christian Kastner}, {and} \bibinfo{person}{Laurie Williams}.} \bibinfo{year}{June 2023}\natexlab{}.
\newblock \showarticletitle{S3C2 Summit 2023-06: Government Secure Supply Chain Summit}.
\newblock \bibinfo{journal}{\emph{https://arxiv.org/abs/2308.06850}} (\bibinfo{year}{June 2023}).
\newblock


\bibitem[{European Union}(2024)]%
        {cra}
\bibfield{author}{\bibinfo{person}{{European Union}}.} \bibinfo{year}{2024}\natexlab{}.
\newblock \bibinfo{title}{{Horizontal cybersecurity requirements for products with digital elements and amending Regulations (EU) No 168/2013 and (EU) No 2019/1020 and Directive (EU) 2020/1828 (Cyber Resilience Act)}}.
\newblock
\newblock


\bibitem[Hamer et~al\mbox{.}(2026)]%
        {hamer2025closing}
\bibfield{author}{\bibinfo{person}{Sivana Hamer}, \bibinfo{person}{Jacob Bowen}, \bibinfo{person}{Md~Nazmul Haque}, \bibinfo{person}{Robert Hines}, \bibinfo{person}{Chris Madden}, {and} \bibinfo{person}{Laurie Williams}.} \bibinfo{year}{2026}\natexlab{}.
\newblock \showarticletitle{{Closing the Chain: How to reduce your risk of being SolarWinds, Log4j, or XZ Utils}}. In \bibinfo{booktitle}{\emph{International Conference on Software Engineering (ICSE)}}.
\newblock


\bibitem[House(2021)]%
        {EO}
\bibfield{author}{\bibinfo{person}{US~White House}.} \bibinfo{year}{May 12, 2021}\natexlab{}.
\newblock \showarticletitle{Executive Order 14028 on Improving the Nation's Cybersecurity}.
\newblock \bibinfo{journal}{\emph{https://www.whitehouse.gov/briefing-room/presidential-actions/2021/05/12/executive-order-on-improving-the-nations-cybersecurity/}} (\bibinfo{year}{May 12, 2021}).
\newblock


\bibitem[Lin et~al\mbox{.}(2025)]%
        {Summit8}
\bibfield{author}{\bibinfo{person}{Elizabeth Lin}, \bibinfo{person}{Jonah Ghebremichael}, \bibinfo{person}{William Enck}, \bibinfo{person}{Yasemin Acar}, \bibinfo{person}{Michel Cucker}, \bibinfo{person}{Alexandros Kapravelos}, \bibinfo{person}{Christian Kastner}, {and} \bibinfo{person}{Laurie Williams}.} \bibinfo{year}{March 2025}\natexlab{}.
\newblock \showarticletitle{S3C2 Summit 2025-03: Industry Secure Supply Chain Summit}.
\newblock \bibinfo{journal}{\emph{https://arxiv.org/pdf/2510.24920}} (\bibinfo{year}{March 2025}).
\newblock


\bibitem[Miller et~al\mbox{.}(2024)]%
        {Summit6}
\bibfield{author}{\bibinfo{person}{Courtney Miller}, \bibinfo{person}{Yasemin Acar}, \bibinfo{person}{Michel Cucker}, \bibinfo{person}{William Enck}, \bibinfo{person}{Christian Kastner}, \bibinfo{person}{Alexandros Kapravelos}, \bibinfo{person}{Dominik Wermke}, {and} \bibinfo{person}{Laurie Williams}.} \bibinfo{year}{August 2024}\natexlab{}.
\newblock \showarticletitle{S3C2 Summit 2024-08: Government Secure Supply Chain Summit}.
\newblock \bibinfo{journal}{\emph{https://arxiv.org/abs/2504.00924}} (\bibinfo{year}{August 2024}).
\newblock


\bibitem[{MITRE}(2025a)]%
        {mitreSHIELD}
\bibfield{author}{\bibinfo{person}{{MITRE}}.} \bibinfo{year}{2025}\natexlab{a}.
\newblock \bibinfo{title}{{Introduction to MITRE Shield}}.
\newblock \bibinfo{howpublished}{\url{https://shield.mitre.org/resources/downloads/Introduction_to_MITRE_Shield.pdf}}.
\newblock


\bibitem[{MITRE}(2025b)]%
        {mitreATLAS}
\bibfield{author}{\bibinfo{person}{{MITRE}}.} \bibinfo{year}{2025}\natexlab{b}.
\newblock \bibinfo{title}{{MITRE ATLAS}}.
\newblock \bibinfo{howpublished}{\url{https://atlas.mitre.org/}}.
\newblock


\bibitem[{MITRE}(2025c)]%
        {mitreATTACK}
\bibfield{author}{\bibinfo{person}{{MITRE}}.} \bibinfo{year}{2025}\natexlab{c}.
\newblock \bibinfo{title}{{MITRE ATT\&CK}}.
\newblock \bibinfo{howpublished}{\url{https://attack.mitre.org/}}.
\newblock


\bibitem[{National Institute of Standards and Technology}(2025)]%
        {nistNVD}
\bibfield{author}{\bibinfo{person}{{National Institute of Standards and Technology}}.} \bibinfo{year}{2025}\natexlab{}.
\newblock \bibinfo{title}{{National Vulnerability Database}}.
\newblock \bibinfo{howpublished}{\url{https://nvd.nist.gov/}}.
\newblock


\bibitem[NIST(2022)]%
        {SSDF}
\bibfield{author}{\bibinfo{person}{NIST}.} \bibinfo{year}{2022}\natexlab{}.
\newblock \showarticletitle{NIST Special Publication 800-218 Secure Software Development Framework (SSDF)}.
\newblock \bibinfo{journal}{\emph{\url{https://nvlpubs.nist.gov/nistpubs/SpecialPublications/NIST.SP.800-218.pdf}}} (\bibinfo{year}{2022}).
\newblock


\bibitem[NTIA(2025)]%
        {NtiA2025}
\bibfield{author}{\bibinfo{person}{NTIA}.} \bibinfo{year}{August 22, 2025}\natexlab{}.
\newblock \showarticletitle{The Minimal Elements of a Software Bill of Materials}.
\newblock \bibinfo{journal}{\emph{\url{https://cisa.gov\/sites\/default\/files\/2025-08\/2025_CISA_SBOM_Minimum_Elements.pdf}}} (\bibinfo{year}{August 22, 2025}).
\newblock


\bibitem[Rahan et~al\mbox{.}(2024)]%
        {Summit7}
\bibfield{author}{\bibinfo{person}{Imranur Rahan}, \bibinfo{person}{Yasemin Acar}, \bibinfo{person}{Michel Cucker}, \bibinfo{person}{William Enck}, \bibinfo{person}{Christian Kastner}, \bibinfo{person}{Alexandros Kapravelos}, \bibinfo{person}{Dominik Wermke}, {and} \bibinfo{person}{Laurie Williams}.} \bibinfo{year}{September 2024}\natexlab{}.
\newblock \showarticletitle{S3C2 Summit 2024-09: Industry Secure Supply Chain Summit}.
\newblock \bibinfo{journal}{\emph{https://arxiv.org/abs/2505.10538}} (\bibinfo{year}{September 2024}).
\newblock


\bibitem[Raymond(1999)]%
        {raymond1999cathedral}
\bibfield{author}{\bibinfo{person}{Eric Raymond}.} \bibinfo{year}{1999}\natexlab{}.
\newblock \showarticletitle{The cathedral and the bazaar}.
\newblock \bibinfo{journal}{\emph{Knowledge, Technology \& Policy}} \bibinfo{volume}{12}, \bibinfo{number}{3} (\bibinfo{year}{1999}), \bibinfo{pages}{23--49}.
\newblock


\bibitem[{The White House}(2025)]%
        {EO14306Ammendment}
\bibfield{author}{\bibinfo{person}{{The White House}}.} \bibinfo{year}{2025}\natexlab{}.
\newblock \bibinfo{title}{{Sustaining Select Efforts To Strengthen The Nation's Cybersecurity and Amending Executive Order 13694 and Executive Order 14144}}.
\newblock \bibinfo{howpublished}{\url{https://www.whitehouse.gov/presidential-actions/2025/06/sustaining-select-efforts-to-strengthen-the-nations-cybersecurity-and-amending-executive-order-13694-and-executive-order-14144/}}.
\newblock


\bibitem[Torres-Arias et~al\mbox{.}(2019)]%
        {torres2019toto}
\bibfield{author}{\bibinfo{person}{Santiago Torres-Arias}, \bibinfo{person}{Hammad Afzali}, \bibinfo{person}{Trishank~Karthik Kuppusamy}, \bibinfo{person}{Reza Curtmola}, {and} \bibinfo{person}{Justin Cappos}.} \bibinfo{year}{2019}\natexlab{}.
\newblock \showarticletitle{in-toto: Providing farm-to-table guarantees for bits and bytes}. In \bibinfo{booktitle}{\emph{28th USENIX Security Symposium (USENIX Security 19)}}. \bibinfo{pages}{1393--1410}.
\newblock


\bibitem[Tran et~al\mbox{.}(2022)]%
        {Summit1}
\bibfield{author}{\bibinfo{person}{Mindy Tran}, \bibinfo{person}{Yasemin Acar}, \bibinfo{person}{Michel Cucker}, \bibinfo{person}{William Enck}, \bibinfo{person}{Alexandros Kapravelos}, \bibinfo{person}{Christian Kastner}, {and} \bibinfo{person}{Laurie Williams}.} \bibinfo{year}{Sept 2022}\natexlab{}.
\newblock \showarticletitle{S3C2 Summit 2022-09: Industry Secure Supply Chain Summit}.
\newblock \bibinfo{journal}{\emph{http://arxiv.org/abs/2307.15642}} (\bibinfo{year}{Sept 2022}).
\newblock


\bibitem[Tystahl et~al\mbox{.}(2024)]%
        {Summit5}
\bibfield{author}{\bibinfo{person}{Greg Tystahl}, \bibinfo{person}{Yasemin Acar}, \bibinfo{person}{Michel Cucker}, \bibinfo{person}{William Enck}, \bibinfo{person}{Christian Kastner}, \bibinfo{person}{Alexandros Kapravelos}, \bibinfo{person}{Dominik Wermke}, {and} \bibinfo{person}{Laurie Williams}.} \bibinfo{year}{March 2024}\natexlab{}.
\newblock \showarticletitle{S3C2 Summit 2024-03: Industry Secure Supply Chain Summit}.
\newblock \bibinfo{journal}{\emph{https://arxiv.org/abs/2405.08762}} (\bibinfo{year}{March 2024}).
\newblock


\bibitem[{White House}(2025)]%
        {EO14144Cybersecurity}
\bibfield{author}{\bibinfo{person}{{White House}}.} \bibinfo{year}{2025}\natexlab{}.
\newblock \showarticletitle{Executive Order 14144 on Strengthening and Promoting Innovation in the Nation's Cybersecurity}.
\newblock  \bibinfo{number}{14144} (\bibinfo{date}{January 16} \bibinfo{year}{2025}).
\newblock
\urldef\tempurl%
\url{https://public-inspection.federalregister.gov/2025-01470.pdf}
\showURL{%
\tempurl}
\newblock
\shownote{Accessed: 2025-04-28}.


\bibitem[Young(2022)]%
        {EO_memo}
\bibfield{author}{\bibinfo{person}{Shalanda~D. Young}.} \bibinfo{year}{2022}\natexlab{}.
\newblock \showarticletitle{M-22-18 Enhancing the Security of the Software Supply Chain through Secure Software Development Practices}.
\newblock \bibinfo{journal}{\emph{https://www.whitehouse.gov/wp-content/uploads/2022/09/M-22-18.pdf}} (\bibinfo{year}{2022}).
\newblock


\bibitem[Zahan et~al\mbox{.}(2023)]%
        {Summit4}
\bibfield{author}{\bibinfo{person}{Nusrat Zahan}, \bibinfo{person}{Yasemin Acar}, \bibinfo{person}{Michel Cucker}, \bibinfo{person}{William Enck}, \bibinfo{person}{Alexandros Kapravelos}, \bibinfo{person}{Christian Kastner}, {and} \bibinfo{person}{Laurie Williams}.} \bibinfo{year}{November 2023}\natexlab{}.
\newblock \showarticletitle{S3C2 Summit 2023-11: Industry Secure Supply Chain Summit}.
\newblock \bibinfo{journal}{\emph{https://arxiv.org/abs/2408.16529}} (\bibinfo{year}{November 2023}).
\newblock


\end{thebibliography}
